\documentclass[aps,prd,showpacs,notitlepage,nofootinbib,superscriptaddress,floatfix,showkeys,twocolumn]{revtex4-1}
\pdfoutput=1
\usepackage{array}
\usepackage[normalem]{ulem} 
\usepackage{verbatim}
\usepackage{graphicx}
\usepackage{amsmath}
\usepackage{amsfonts}
\usepackage{amssymb}
\usepackage{empheq}
\usepackage{xcolor, soul}
\usepackage{epstopdf}
\usepackage{float}
\usepackage{subfigure}
\usepackage{soul}
\usepackage{hyperref}
\hypersetup{
     colorlinks   = true,
     citecolor    = red,
     linkcolor    = blue,
     urlcolor     = blue,
}
\newcommand{\beal}{\begin{aligned}}
\newcommand{\eeal}{\end{aligned}}
\newcommand{\bea}{\begin{eqnarray}}
\newcommand{\eea}{\end{eqnarray}}
\newcommand{\red}{\color{red}}
\newcommand{\blue}{\color{blue}}
\newcommand{\noi}{\noindent}
\def\beq{\begin{equation}}
\def\eeq{\end{equation}}
\def\be{\begin{equation}}
\def\ee{\end{equation}}

\def\bse{\begin{subequations}}
\def\ese{\end{subequations}}

\def\Min{M_{\rm in}}
\def\Hin{H_{\rm in}}
\def\Tin{T_{\rm in}}
\def\tin{t_{\rm in}}
\def\ain{a_{\rm in}}

\def\He{H_{\rm end}}

\newcommand{\Tbh}{T_{\rm BH}}
\newcommand{\Mbh}{M_{\rm BH}}
\newcommand{\xev}{x_{\rm ev}}
\newcommand{\aev}{a_{\rm ev}}
\newcommand{\tev}{t_{\rm ev}}

\newcommand{\arh}{a_{\rm RH}}
\newcommand{\Trh}{T_{\rm RH}}
\newcommand{\rhorh}{\rho_{\rm RH}}

\graphicspath{{./figs/}}
\begin{document}

\title{Primordial Black Holes in a Thermal Bath: from Absorption to Evaporation}

\title{When Primordial Black Holes Absorb During the Early Universe}

\author{Md Riajul Haque}%
\email{riaj.0009@gmail.com}
\affiliation{\,Tsung-Dao Lee Institute \& School of Physics and Astronomy, Shanghai Jiao Tong University,
Shanghai 201210, China}.
\author{Rajesh Karmakar}
\email{rajesh@shu.edu.cn}
\affiliation{Department of Physics, Shanghai University, 99 Shangda Road, Shanghai, 200444, China}
\author{Yann Mambrini}
\email{yann.mambrini@ijclab.in2p3.fr}
\affiliation{
	Universit\'e Paris-Saclay, CNRS/IN2P3, IJCLab, 91405 Orsay, France 
}

\date{\today}

\begin{abstract}
We study the evolution of primordial black holes (PBHs) formed in the early universe in the presence of a surrounding thermal bath. By incorporating the effects of thermal 
absorption, we show that PBHs can undergo significant mass growth, leading to extended lifetimes and substantial deviations from the standard Hawking evaporation scenario. 
We find a critical collapse efficiency, $\gamma_{\rm c} \simeq 0.395$, above which the PBH mass grows without bound. 
This correction has profound implications for both PBH-induced reheating and dark matter (DM) production. Specifically, we find that the 
reheating temperature can be suppressed, and the DM parameter space for the PBH 
reheating scenario can undergo $\mathcal{O}(10)$–$\mathcal{O}(10^4)$ corrections, depending on the PBH formation mass and collapse efficiency. Moreover, our results significantly shift the 
parameter space in which PBHs can account for the entirety of the DM. To the best of 
our knowledge, this is the first comprehensive phenomenological study to incorporate 
thermal absorption into PBH evolution and quantify its impact on cosmological observables.

\end{abstract}
\maketitle
\section{Introduction}
In the early universe, initial inhomogeneities in matter density could have given rise 
to highly compressed regions. These overdense areas, enhanced by primordial density fluctuations, are thought to have undergone gravitational collapse, possibly forming 
primordial black holes (PBHs) \cite{Carr:1974nx}. In contrast to stellar BHs, 
PBHs could have been formed with a very broad mass spectrum, 
depending on the scale and amplitude of the 
initial over densities \cite{Carr:1975qj}, which have fascinating implications for 
cosmology.

Ultralight PBHs with masses below $10^9\,\mathrm{g}$ can simultaneously reheat the universe~\cite{Hidalgo:2011fj, Martin:2019nuw, Hooper:2019gtx, Hooper:2020evu, Hooper:2020otu, Bernal:2020bjf, RiajulHaque:2023cqe, Barman:2024slw, Sanchis:2025awq}, account for the present DM abundance~\cite{Morrison:2018xla, Gondolo:2020uqv, Bernal:2020bjf, Green:1999yh, Khlopov:2004tn, Dai:2009hx, Allahverdi:2017sks, Lennon:2017tqq, Masina:2020xhk, Baldes:2020nuv, Hooper:2019gtx, Bhaumik:2022zdd, Cheek:2021odj, Cheek:2021cfe, Bernal:2022oha, Borah:2022iym, Haque:2024eyh, Haque:2024cdh}, and generate observable stochastic GW signals~\cite{Inomata:2019zqy, Domenech:2021ztg, Papanikolaou:2020qtd, Gross:2024wkl, Bhaumik:2024qzd, Paul:2025kdd, Domenech:2024wao}, motivating ongoing and future GW searches~\cite{TitoDAgnolo:2024uku, Domcke:2024eti, Maity:2024cpq}. Stable PBHs in the mass range $10^{17} - 10^{21}\,\mathrm{g}$ can also account for the full DM abundance~\cite{Villanueva-Domingo:2021spv,LISACosmologyWorkingGroup:2023njw}, with additional GW signals from mergers potentially explaining LIGO/Virgo events~\cite{LIGOScientific:2018mvr} and being testable by LISA and DECIGO in the intermediate and supermassive mass ranges~\cite{Seto:2001qf, Kawamura:2020pcg, Carr:2020gox, Carr:2020xqk, Green:2020jor}. PBHs have further been proposed as seeds for supermassive BHs and sources of gravitational lensing~\cite{Wang:2021lij}. Formed in the early universe, they offer a unique framework for probing both fundamental black hole physics~\cite{Carr:2014eya, Auffinger:2022khh} and cosmological phenomenology.


Classically, BHs are considered perfect black bodies because they absorb all matter 
and radiation that crosses the event horizon \cite{Misner:1973prb}. In the early stages 
of cosmic evolution, when the Universe was dominated by radiation and matter density, 
this process became particularly significant for PBHs. The ``catastrophic" 
consequence of radiation absorption, leading to an unbounded growth of BH, was first identified by Zeldovich and Novikov in \cite{Zeldovich:1967lct}. In a similar manner, the impact of accretion onto PBHs in the presence of a background field
have been studied, including the cosmic expansion
\cite{Carr:2010wk, Harada:2004pf,Das:2025vts}. On the other hand, the semiclassical prediction of Hawking radiation implies that a BH of mass $M_{\rm BH}$ emits thermal 
radiation and gradually evaporates, 
acquiring a characteristic temperature $T_{\rm BH}=\frac{M_P^2}{M_{\rm BH}}$,
where $M_P = 1/\sqrt{8\pi G} \simeq 2.4 \times 10^{18} \, \rm{GeV}$ is the reduced Planck mass \cite{Hawking:1975vcx}. As the background temperature was extremely high in the early 
universe, the competition between the evaporation process with the thermal-absorption prompts for closer look. 

Indeed, a recent analysis suggested that the evolution of the BHs depends critically on the temperature of the surrounding thermal 
bath \cite{Barrau:2022bfg}. A very similar conclusion was also reached for BHs in 2D JT gravity, placed in a reservoir \cite{Chen:2020jvn} at finite temperature,
or taking into account thermal correction to the occupation number \cite{Kalita:2025foa}. However, none of these processes capture the significant mass 
growth due to {\it direct absorption} of thermal radiations \cite{Zeldovich:1967lct}. Therefore,
these studies motivate the investigation of PBHs in the radiation-dominated era of the early universe.

In the present work, we analyse the system of Schwarzschild PBHs placed within the primordial plasma\footnote{We leave for future study the extension to Kerr spacetime.}.  Without favoring any particular accretion 
model, we include all the prevalent degrees of freedom existing during the phase of radiation domination, with their respective absorption cross section. In 
particular, we study how the relative temperature of the BH compared to the 
plasma temperature influences the absorption process and thereby the evolution of the PBHs {\footnote{In the direct absorption process, we also assume that the particle mean free path is greater than the black hole horizon radius \cite{Zeldovich:1967lct}. Although PBHs form with a horizon radius of the order of the Hubble scale, even when interaction rates exceed the Hubble rate, the preceding condition might be satisfied as the universe evolve \cite{Kolb:1990vq}.}}. With regard to previous analysis, we show that the unbounded growth of PBHs can be circumvented 
or at least can be brought down to saturation in a finite time, 
due to the cosmological evolution of the Universe. 
Furthermore, we investigate the consequence of PBH mass growth in the context 
of PBH reheating
\cite{RiajulHaque:2023cqe}
and how it affects the DM abundance produced by PBHs evaporation \cite{Haque:2023awl}.

The paper is organized as follows. In Sec.~\ref{bh.abs}, we briefly discuss the absorption cross section of BH, distinguishing the low and high frequency regimes of the absorbed  particles. We then look at the impact of the absorption on PBH lifetime in Sec.~\ref{Sec:evolution}, where we compute the condition for which the PBH mass saturates after growing in a finite duration. 
We then look at the consequences on the reheating and DM production in Sec.~\ref{Sec:reheating}, before concluding in Sec.~\ref{Sec:conclusion}.
Throughout our analysis, we use the natural unit system, $\hbar=c=k_B=1$, whereas we  kept the gravitational constant, $G=\frac{1}{8 \pi M_{P}^2}$ as it is, unless otherwise specified. 

\section{
Black hole formation, absorption and evaporation: a brief overview
}\label{bh.abs}

\subsection{Formation}

We assume that primordial black holes form during a radiation-dominated (RD) era as a result of the gravitational collapse of density fluctuations. Consequently, the formation mass of the PBHs, denoted by $\Min$, can be expressed as

\be
M_{\rm in}=\gamma \rho_{\rm R}(t_{\rm in})\frac{4}{3}\pi\frac{1}{H_{\rm in}^3}=4\pi\gamma\frac{M^2_P}{H_{\rm in}}
\simeq 1.3~\gamma ~{\rm g}\left(\frac{10^{14}~\rm GeV}{\Hin}\right)
\,,
\label{Eq:min}
\ee

\noi
where $\gamma$ is the collapse efficiency parameter\footnote{It is worth noting that more accurate methods exist for estimating the formation mass of primordial black holes (PBHs), which take into account both the nature of the cosmological background and the specific profile of scalar perturbations~\cite{Musco:2012au, Musco:2008hv, Hawke:2002rf, Niemeyer:1997mt, Escriva:2021pmf, Escriva:2019nsa, Escriva:2020tak, Escriva:2021aeh}. However, a fully analytical understanding of the values of $\gamma$, particularly in the context of PBH formation during the post-inflationary epoch is still lacking.}.
Here, $H_{\rm in}$ denotes the Hubble parameter at the time of PBH formation, which is related to the background radiation temperature via

\be
H_{\rm in} = \sqrt{\frac{\rho_R (t_{\rm in})}{3M_{\rm p}^2}} = 
\sqrt{\frac{\alpha}{3}}\,\frac{T_{\rm in}^2}{M_{\rm p}},
\label{Eq:hin}
\ee

\noi
with $\alpha=g_{\ast}(T)\frac{\pi^2}{30}$ and $g_{\ast}(T)$ being the effective number of relativistic degrees of freedom at temperature $T$.
$T_{\rm in}$ is the radiation temperature at the time of formation. 
Note that, from the hypothesis $\Hin\lesssim \He$, $\He$ being the Hubble rate at the end of inflation $\He \lesssim 10^{14}$ GeV \cite{Planck:2018jri}, Eq.~(\ref{Eq:min}) gives $\Min\gtrsim 1$ g for $\gamma$
of the order one.

Combining Eqs.~(\ref{Eq:min}) and (\ref{Eq:hin}), we obtain

\beq
\Tin = \left(\frac{4 \pi \gamma}{\sqrt{\alpha/3}}\right)^\frac12\frac{M_P^\frac32}{\Min^\frac12}\simeq
9.7\times 10^{15}\sqrt{\gamma}\left(\frac{1~\rm g}{\Min}\right)^\frac12~\rm GeV\,.
\label{Eq:tin}
\eeq

\subsection{Absorption}

Naively, in the classical picture the event horizon of a BH acts as a one-way 
membrane, absorbing anything that crosses it. Moreover, the spacetime geometry around a BH gives rise to an effective potential barrier, whose features depend on the spin 
of the incoming particles and other relevant parameters. Consequently, the fate of an incoming particle, whether it is absorbed or scattered, depends on these parameters, 
leading to a frequency-dependent absorption probability. For the purposes of later discussion, we will particularly focus on the asymptotic behaviour of the absorption 
cross section in different frequency regimes.\\
\begin{center}
\textit{High-frequency absorption cross section}
\end{center}

In the regime where the frequency $\omega$ of an incoming particle greatly exceeds the characteristic scale of the black hole (i.e., $\omega \gg 1/r_s$, where

\beq
r_s = 2GM_{\rm BH}=\frac{M_{\rm BH}}{4 \pi M_P^2}\,,
\label{Eq:rs}
\eeq

\noi
is the Schwarzschild radius), the absorption cross section asymptotically approaches the geometric optics limit. This limit can be derived by analyzing the classical capture cross section for null geodesics, given by $\sigma_{\rm hf} = \pi b_c^2$, where $b_c = 3\sqrt{3}\,GM_{\rm BH}$ is the critical impact parameter in a Schwarzschild background \cite{Mambrini:2021cwd,Padmanabhan:2010zzb}. This yields:
\be
\sigma_{\rm hf} = 27\pi G^2 M_{\rm BH}^2=\frac{27}{64 \pi}\frac{M_{\rm BH}^2}{M_P^4} \,,
\label{Eq: high frequency}
\ee

\noi
which is commonly referred to as the high-frequency or optical limit of the absorption cross section. This result is universal and applies to both bosonic and fermionic fields, provided the frequency is sufficiently high~\cite{Crispino:2007qw}.\\

\noindent

\begin{center}
\textit{Low-frequency absorption cross section}
\end{center}

 In the opposite regime, where the thermal wavelength is larger than the 
 Schwarzschild radius (i.e., $\omega \ll 1/r_s$), the absorption cross 
 section is suppressed and becomes sensitive to the spin $s$ of the incoming 
 particle. For relativistic particles in this low-frequency limit, the cross 
 sections take the following approximate forms \cite{MacGibbon:1990zk} (see the Appendix for a brief discussion on the derivation):
\be \label{Eq: low frequency}
\sigma_{\rm lf}\sim
\left\{
\beal
&16\pi G^2M_{\rm BH}^2 =\frac{M_{\rm BH}^2}{4 \pi M_P^4},~~~~~~~s=0\,,\\
&2\pi G^2M_{\rm BH}^2=\frac{M_{\rm BH}^2}{32 \pi M_P^4},~~~~~~~~~~s=1/2\,,\\
&\frac{64\pi}{3}G^4M_{\rm BH}^4\,\omega^2 =\frac{M_{\rm BH}^4}{192 \pi^3 M_P^8\,\omega^2},~~~~~s=1\,.\\
\eeal\right.
\ee

For spin-1 particles (e.g., photons), the absorption cross section becomes 
negligibly small in the $\omega \to 0$ limit, whereas spin-$1/2$ fermions and 
scalar particles (spin-0) have significantly higher absorption probabilities 
in this regime. These results allow us to estimate the energy absorbed by a 
BH placed in a relativistic thermal bath by integrating over the equilibrium 
energy density of the bath, weighted by the relevant absorption cross section 
in either frequency regime. 

 The mass evolution of the black hole
 due to the 
 absorption of a local plasma at temperature $T$ and density of energy 
 $\rho_R(T)$ is then described by 
\be
\frac{dM_{\rm BH}}{dt} =
\sigma_i \rho_R 
=\delta_i \frac{T^4}{\Tbh^2}\,,
\label{Eq:absorption}
\ee

\noi
with

\beq
\Tbh=\frac{M_P^2}{\Mbh}\,,
\eeq

\noi
and 


\beq
\delta_{\rm hf}=\frac{9\pi}{640}g_*\,,
\eeq

\beq
\delta_{\rm lf}\sim
\left\{
\beal
&\frac{1}{30}g_*\,,~~~~~~~s=0\,,\\
&\frac{1}{240} g_*,~~~~~~~~~~s=1/2\,,\\
&\frac{1}{1440 \pi^2}\frac{\omega^2}{\Tbh^2}g_*,~~~~~s=1\,.\\
\eeal\right.
\eeq

\subsection{Evaporation}

To accurately track the evolution of PBHs, it is essential to also account for Hawking evaporation given by 

\be
\frac{d \Mbh}{dt} = -\epsilon \frac{M_{\rm p}^4}{\Mbh^2}=-\epsilon \Tbh^2\,,
\label{Eq:evaporation}
\ee
where $\epsilon \equiv \frac{27}{4} \left( \frac{\pi}{480} \right) g_{\ast}(T_{\rm BH})$ characterizes the evaporation rate. The factor $\frac{27}{4}$ accounts for the greybody factor \cite{Arbey:2019mbc, Cheek:2021odj, Baldes:2020nuv}. Notably, the above expression is obtained by approximating the absorption cross section for all the degrees of freedom in the geometric optics limit, with $g_{\ast}(T_{\rm BH})$ representing the number of relativistic degrees of freedom at the black hole temperature $T_{\rm BH}$ \cite{Baldes:2020nuv, Cheek:2021odj}.
We can now combine
Eqs.~(\ref{Eq:absorption}) and (\ref{Eq:evaporation}) to obtain the key equation

\beq
\frac{d \Mbh}{dt}=\delta_i\frac{T^4}{\Tbh^2}\left[1-\frac{\epsilon}{\delta_i}\left(\frac{\Tbh}{T}\right)^4\right]\,.
\label{Eq:masterequation}
\eeq

 Note that in the high-frequency (geometric-optics) limit, the absorption cross section approaches the classical capture cross section, determined by null-ray capture by the horizon. 
By time-reversal invariance of the linear wave equation and detailed balance, the same transmission probability governs Hawking emission, as a result $\epsilon=\delta_{hf}$. 
Consequently, when the ambient temperature equals the black-hole temperature, $T = T_{\rm BH}$, absorption and evaporation exactly balance, yielding zero net energy flux.

Eq.~(\ref{Eq:masterequation}) deserves some comments. First of all, we clearly distinguish two regimes, $T\gg \Tbh$, and $T\ll \Tbh$. In the first
regime, the absorption dominates the evolution process, whereas in the second phase, the evaporation dominates the PBH behavior. Note also that $\Tbh$
is not a {\it thermal} temperature stricto-sensu, but the Hawking temperature, which particles respecting a thermal distribution. 
Second, one should also consider $T \gg \frac{2 \pi}{r_s}$ and $T \ll \frac{2 \pi}{r_s}$
to distinguish the high frequency and low frequency absorption modes.
One then to check the condition for an effciient absorption

\section{PBH evolution}
\label{Sec:evolution}

\subsection{Condition for absorption}

For the absorption to be effective, one needs to check that at formation time, $\tin$, corresponding to the scale factor $\ain$, $\Tin>\Tbh$.
For that, we rewrite Eq.~(\ref{Eq:masterequation}) as function of the scale factor $a$ using $\frac{d}{dt}=Ha\frac{d}{da}$

\bea
&&
\frac{d\Mbh}{da}
=\frac{3 \delta_i M_P^2 H}{\alpha a \Tbh^2}\left[1-\frac{\epsilon}{\delta_i}\left(\frac{\Tbh}{T}\right)^4\right]
\nonumber
\\
&&
=\delta_i\sqrt{\frac{3}{\alpha}} \frac{M_P}{a}\frac{T^2}{\Tbh^2}
\left[1-\frac{\epsilon}{\delta_i}\left(\frac{\Tbh}{T}\right)^4\right]
\label{Eq:masterequationbis}
\\
&&
=\frac{12 \pi \gamma \delta_i}{\alpha}\frac{\Mbh^2}{\Min}
\frac{\ain^2}{a^3}\left[
1-\frac{\alpha \epsilon}{48 \pi^2 \gamma^2\delta_i}\frac{M_P^2 \Min^2}{\Mbh^4}\left(\frac{a}{\ain}\right)^4
\right]
\nonumber
\,,
\eea

\noi
or

\beq
\boxed{
\frac{d\Mbh}{dx}
=\frac{12 \pi \gamma \delta_i}{\alpha}\frac{\Mbh^2}{\Min}
\frac{1}{x^3}\left[
1-\frac{\alpha \epsilon}{48 \pi^2 \gamma^2\delta_i}\frac{M_P^2 \Min^2}{\Mbh^4}x^4
\right]
\label{Eq:masterequationter}
 \,,
 }
\eeq

\noi
which can be written

\beq
\boxed{
\frac{dR}{dx}
=\frac{12 \pi \gamma \delta_i}{\alpha} R^2
\frac{1}{x^3}\left[
1-\frac{\alpha \epsilon}{48 \pi^2 \gamma^2\delta_i}\frac{M_P^2}{\Min^2}\frac{x^4}{R^4}
\right]
\label{Eq:masterequationR}
 \,,
 }
\eeq

\noi
with $R(x)=\frac{M(x)}{\Min}$, $x=a/\ain$, and where we supposed a radiation dominated Universe. The condition for an efficient absorption at $\ain$ can be read from Eq.~(\ref{Eq:masterequationter}), asking for the absorption term to dominate at $x= 1$, or 

\beq
\Min\gtrsim  \sqrt{\frac{\alpha \epsilon}{3 \delta_i}} \frac{M_P}{4 \pi \gamma} \simeq \frac{0.12}{\gamma}~\sqrt{\frac{\epsilon}{\delta_i}}~\times 10^{-5}~\rm g \,,
\eeq

\noi
corresponding to, using Eq.~(\ref{Eq:min})

\beq
\Hin \lesssim 16 \pi^2 \gamma^2 \sqrt{\frac{3 \delta_i}{\alpha \epsilon}}M_P\simeq 1.6 \left(\frac{\gamma}{0.12}\right)^2\sqrt{\frac{\delta_i}{\epsilon}}\times 10^{18}~\rm GeV\,,
\eeq

\noi
which is always the case considering $\He \lesssim 10^{14}$ GeV \cite{Planck:2018jri}.


\begin{figure}[!ht]
\centering
\includegraphics[width = 0.45\textwidth]{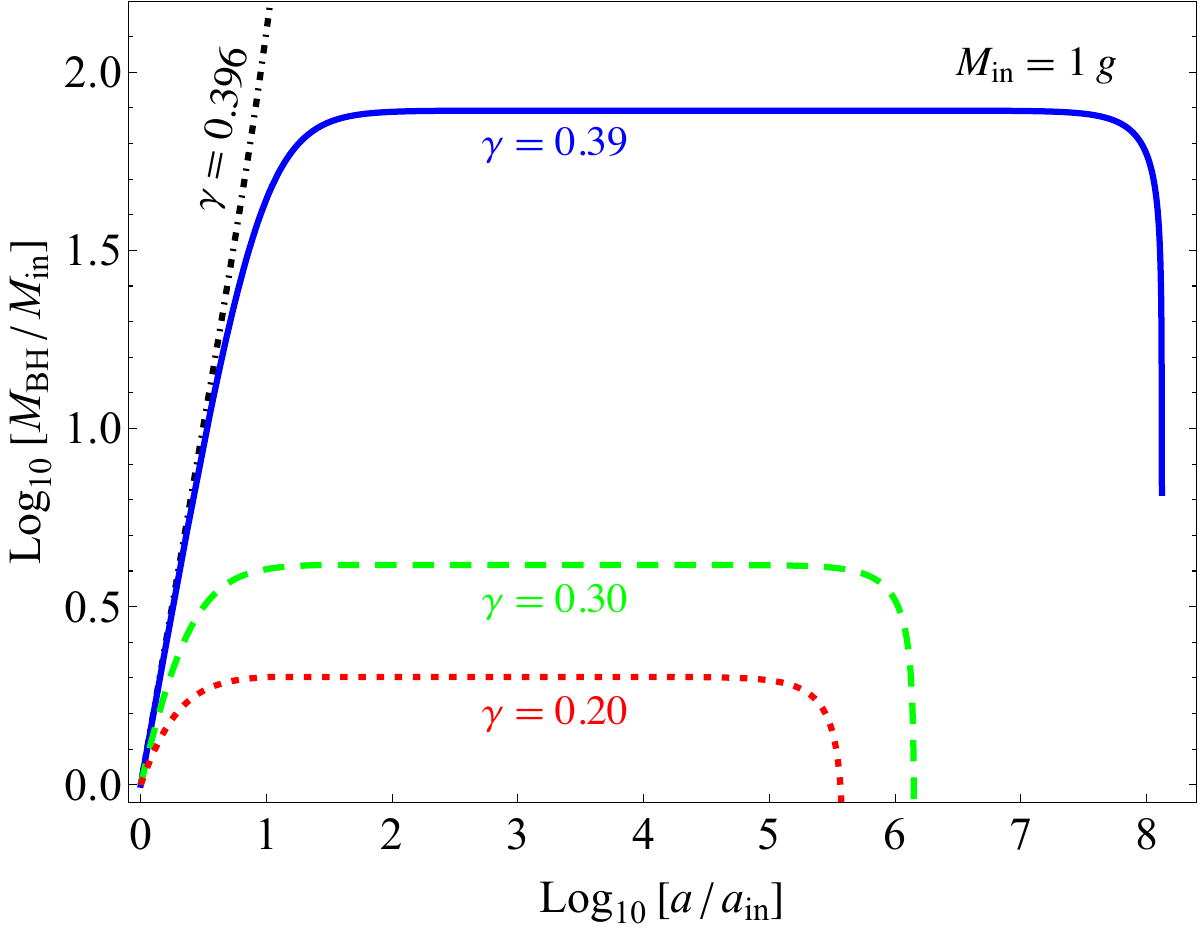}
\caption{Evolution of the PBH mass, normalized to its formation mass $\Min$,
for $\Min=1$g, as function of the scale factor relative to the formation time, $\frac{a}{\ain}$, for four different values of the parameter $\gamma$, which denotes the ratio of the horizon mass to the PBH mass at formation, see Eq.~(\ref{Eq:min}). The red dotted line represents the standard case with $\gamma = 0.2$, while the black dot-dashed line shows the diverging solution for $\gamma=0.396$, values slightly above the critical threshold, $\gamma_{\rm c}=32/81\sim 0.395$, see Eq.(\ref{Eq:deltac}).}
\label{Fig:massevolution}
\end{figure}

We show in Fig.~\ref{Fig:massevolution} the evolution of the PBH mass as function of $x=a/\ain$, for $\Min=1$ g and four different values of $\gamma= 0.2, 0.3, 0.39$ and $0.396$. Whereas $\gamma=0.2$ represents 
the standard value for $\gamma$ within a radiative dominated Universe
($\gamma=(\frac13)^\frac32\simeq 0.2$) \cite{Carr:1974nx}, leading to an increase of the mass $\Mbh\simeq 2\, \Min$ before the evaporation, this changes drastically for increasing values of $\gamma$. Indeed, for $\gamma=0.39$, the mass increases up to $\Mbh\simeq 78 \,\Min$, even diverging for a critical value $\gamma > 0.395$.

\subsection{Effect on the mas evolution}

The evolution of $\Mbh(a)$, or equivalently $R(x)$ in Fig.~\ref{Fig:massevolution} is easy to understand. The BH
first increases its mass by absorbing the radiation which surrounds it. Then, it saturates due to the expansion at a value given by the limit $a\rightarrow \infty$ of Eq.~(\ref{Eq:masterequationR}), where we keep only the first term, the absorption one. Neglecting the evaporation process, integrating  Eq.~(\ref{Eq:masterequationR}), we obtain

\beq
R(x)=\frac{\Mbh(x)}{\Min}=\frac{1}{1-\frac{6 \pi \gamma \delta_i}{\alpha}(1-\frac{1}{x^2})} \,,
\eeq

\noi
or, for $x\gg 1$,

\beq
\left.R(x)\right|_{x\gg 1}=R_\gamma\simeq \frac{1}{1-\frac{6 \pi \gamma \delta_i}{\alpha}}\,,
\label{Eq:assymptoticr}
\eeq
\noi
which gives
$\left.R(x)\right|_{x \gg 1}\simeq 4$ for
$\gamma = 0.3$ and $\delta_i=\delta_{\rm hf}$, corresponding indeed to our numerical result observed in Fig.~\ref{Fig:massevolution}. Note also that the evolution of $R(x)$ during the absorption phase, does not depend on the initial mass $\Min$. We show in table (\ref{Tab:rgamma}) the value of $R_\gamma$ for
different values of $\gamma$.

\begin{table}[h]
    \centering
    \begin{tabular}{|>{\centering\arraybackslash}m{3cm}|>{\centering\arraybackslash}m{2cm}|}
    \hline
     $\gamma$   & $R_\gamma$  \\
     \hline
     \hline
       0  & 1\\
       \hline
       0.1 & 1.34\\
       \hline
       0.2 & 2.03\\
       \hline
       0.3 & 4.16\\
       \hline
        0.39 & 78.05\\
       \hline
       $32/81\simeq 0.395$ & $\infty$\\
       \hline
    \end{tabular}
    \caption{Value of $R_\gamma=\frac{\Mbh}{\Min}$ definded in Eq.~(\ref{Eq:assymptoticr}) for different values of 
    $\gamma$. Note that $\gamma\simeq 0.395$ should be treated as the point of unbounded growth, see Eq.~(\ref{Eq:deltac}).}
    \label{Tab:rgamma}
\end{table}

There is also a value for the $\gamma$ parameter 
beyond which absorption continues without being slowed down by the expansion rate, as one can see in Fig~\ref{Fig:massevolution}. 
This critical value for $\gamma=\gamma_c$ corresponds to
the situation 

\beq
\frac{6 \pi \gamma \delta_i}{\alpha}>1\,,
\eeq

\noi
or

\beq
\gamma \gtrsim  \gamma_c=\frac{\alpha}{6 \pi\delta_i}=\frac{32}{81}\simeq 0.395\,,
\label{Eq:deltac}
\eeq

\noi
for $\delta_i=\delta_{\rm hf}$. This is indeed for this critical value that we observe a divergent behavior for $\Mbh(x)$ in Fig.~\ref{Fig:massevolution}. This effect is one important result of our work. We note that, there exists a critical value for the PBH collapse efficiency above which the absorption is so efficient that the radiation surrounding the PBH horizon is continuously absorbed, without being stopped by the expansion rate.

\subsection{Effect on the evaporation}

As one can see from Eq.~(\ref{Eq:masterequationbis}), there exist a value of $x=a/\ain$ above which the evaporation begins to be efficient. This corresponds to

\beq
x=\left(\frac{48 \pi^2 \gamma^2 \delta_i}{\alpha \epsilon}\right)^\frac14\sqrt{\frac{\Min}{M_P}}R(x)\,,
\eeq

\noi
when the derivative in Eq.~(\ref{Eq:masterequationter}) becomes  negative. However, even though evaporation begins to dominate the evolution of PBH, its effects are only effective much later, when $d\Mbh\sim \Mbh$. Solving Eq.~(\ref{Eq:masterequationter}), neglecting the absorption part, and substituting $\Min \rightarrow R_\gamma \times \Min$, we
obtain

\beq
\xev = \frac{\aev}{\ain} \simeq 
\sqrt{\frac{8 \pi \gamma}{3 \epsilon}}\, R_\gamma^\frac32\, \frac{\Min}{M_P}
\,,
\label{Eq:xev}
\eeq

\noi
or, considering the assymptotical value of $R_\gamma$ given by Eq.~(\ref{Eq:assymptoticr}) for $\gamma=0.3$,


\beq
\xev^{\gamma=0.3}\simeq 1.4\times 10^6\,,
\eeq

\noi
and


\beq
\xev^{\gamma=0.2}\simeq 4\times 10^5\,,
\eeq

\noi
for $\Mbh=1$g, which is effectively what we observe in Fig.~\ref{Fig:massevolution}.
We can also express our result as function of time, using

\bea
&&
\tev=\frac{1}{2 H(\aev)}=\frac{1}{2 \Hin}\left(\frac{\aev}{\ain}\right)^2
=\frac{\xev^2}{2\Hin}
\nonumber
\\
&&
=\frac{\xev^2 \Min}{8 \pi \gamma M_P^2} =\frac{R_\gamma^3}{3 \epsilon}\frac{\Min^3}{M_P^4}\,,
\label{Eq:tev}
\eea

\noi
where we used Eq.~(\ref{Eq:min}), or


\beq
\tev \simeq 1.8\times 10^{-26}~{\rm s.}\,, 
\eeq

\noi
for $\gamma=0.3$ and $\Min=1$ g. In general, it is always possible to convert from scale factor to time using the relation (\ref{Eq:tev}), or

\beq
t=\frac{x^2}{8 \pi \gamma}\frac{\Min}{M_P^2}\,.
\label{Eq:t=f(x)}
\eeq

\subsection{Dependence on the initial mass $\Min$}

\begin{figure}[!ht]
\centering
\includegraphics[width = 0.45\textwidth]{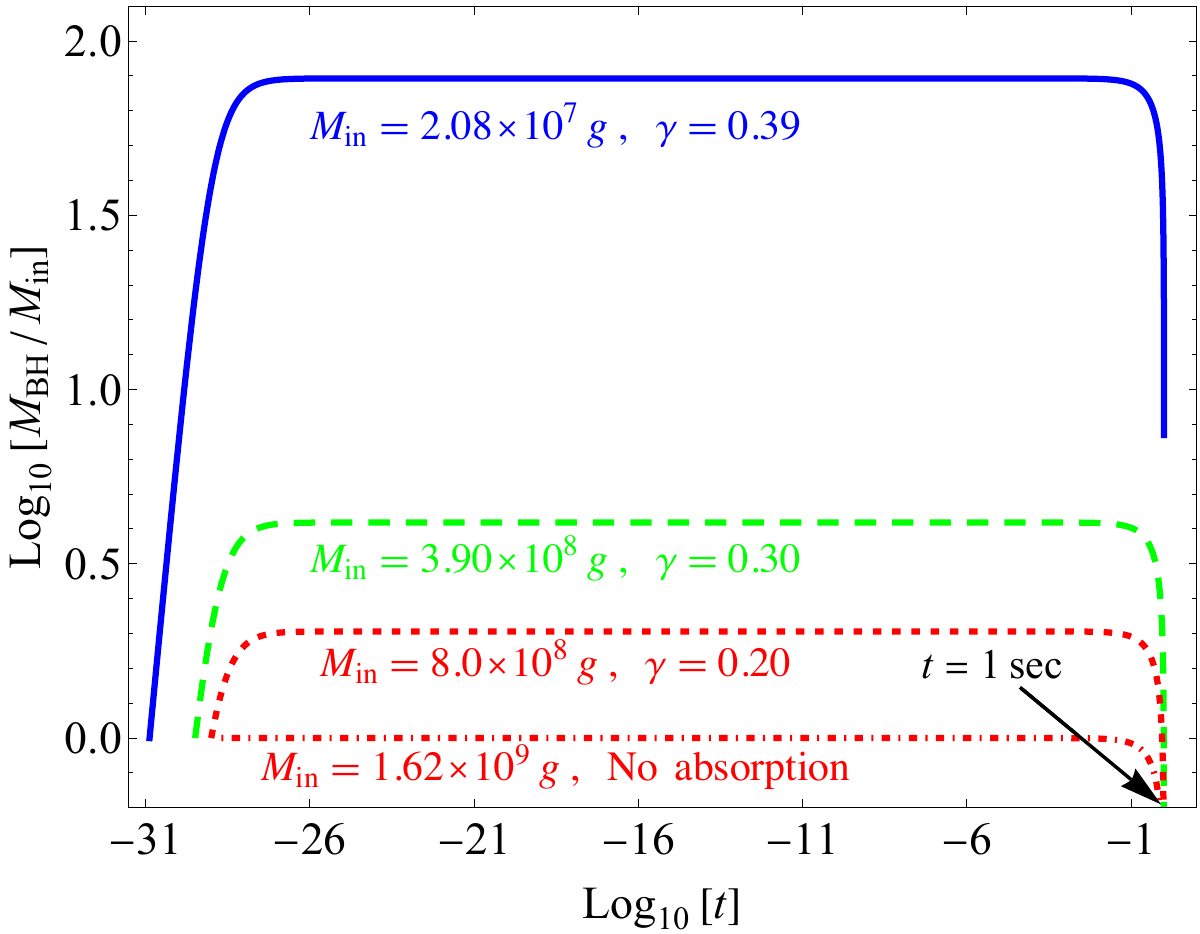}
\caption{Evolution of the PBH mass, normalized to its formation mass, as a function of time for different values of $\gamma$ is shown. The blue solid, gree dashed, and red dotted lines correspond to cases where radiation absorption is included, while the red dot-dashed line represent scenarios where absorption is neglected or $\gamma$ is chosen such a way that absorption has no effect. In all plots, the PBH formation mass are selected so that evaporation occurs around the BBN timescale, i.e., 1 second.}
\label{Fig:evaporation}
\end{figure}

\begin{figure}[!ht]
\centering
\includegraphics[width = 0.45\textwidth]{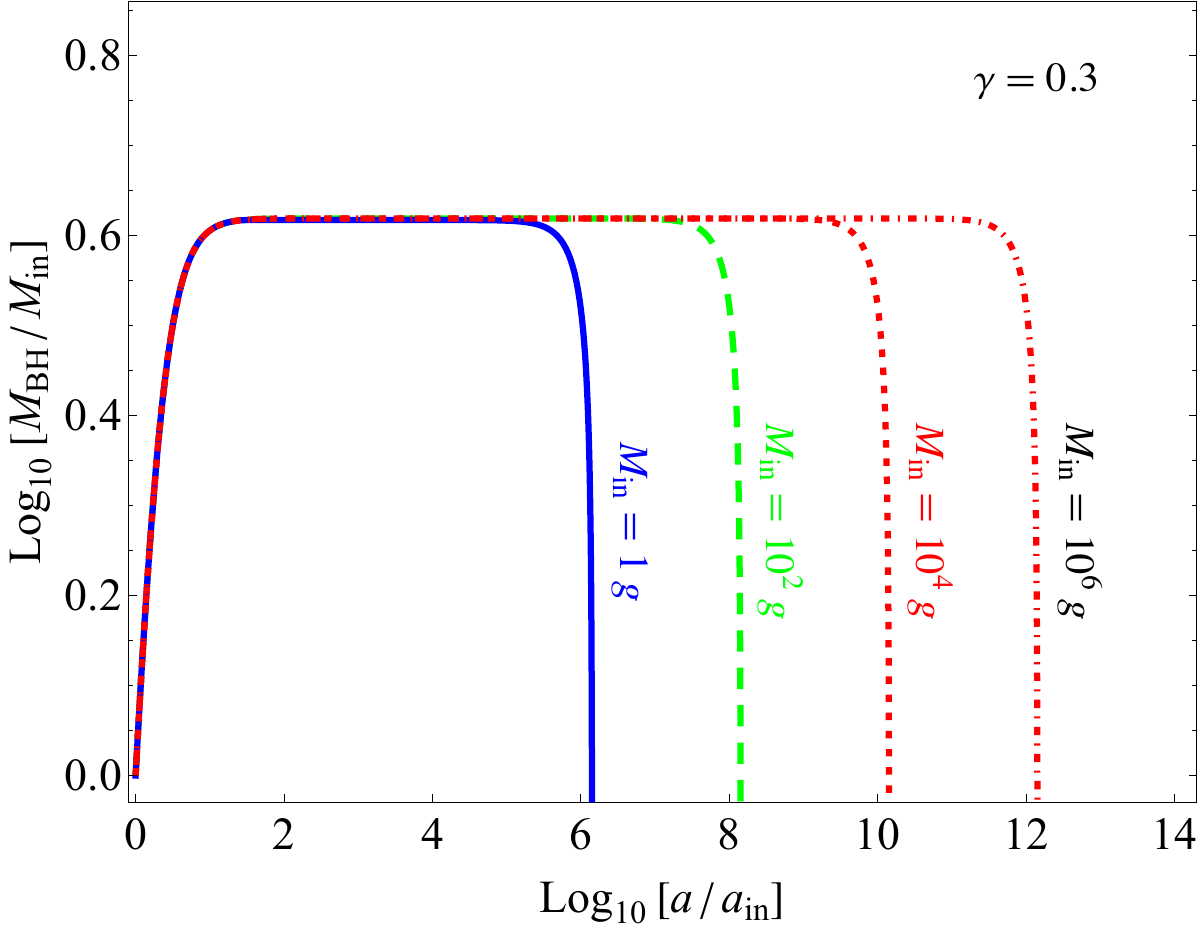}
\caption{Evolution of the PBH mass, normalized to its formation mass $\Min$,
for different values of $\Min$ (1g, $10^2$g, $10^4$g and $10^6$g), as function of the 
scale factor relative to the formation time, $\frac{a}{\ain}$.}
\label{Fig:m(a)}
\end{figure}

We can also ask for which value of $\Min$ the evaporation occurs before 1 second, taking into account the absorption process. 
The result is shown in 
Fig.~\ref{Fig:evaporation}, where we adapted the initial PBH mass $\Min$, fixed to
$\tev=1$ second, for different values of $\gamma$, to ensure a PBH decay before the beginning of the BBN era. We see that, whereas without taking into account the absorption process, a PBH of mass $\Min\simeq 10^9$ grams was able to evaporate before BBN time, for $\gamma=0.3$ one needs to lower this bound on $\Min$ by almost a factor $4$ due to the increase of $\Mbh$ during the absorbing phase. More generally, we can study the evolution of a primordial black hole for 
different values of $\Min$, and $\gamma=0.3$. In Fig.~\ref{Fig:m(a)} we 
present the evolution $\Mbh(a)$ for $\gamma=0.3$ and different value of $\Min$ (1g, $10^2$g, $10^4$g, and $10^6$g). We clearly see that the increase of the 
mass due to the absorption is the same, whatever is the value of $\Min$.
However, the evaporation time decreases for increasing values of $\Min$,
respecting the condition (\ref{Eq:xev}).
We also reexamine the lower bound on the PBH mass required for the black holes to survive until the present epoch. This bound is obtained by setting the PBH evaporation time equal to the current age of the Universe,
\( t_0 \simeq 4.35 \times 10^{17} \,\mathrm{s} \).
In the conventional treatment, where absorption effects are neglected, this requirement leads to $M_{\rm in} \gtrsim 1.2 \times 10^{15}\ {\rm g}$. Once absorption from the ambient thermal plasma is incorporated, the stability bound is modified to $M_{\rm in} \gtrsim \left(2.9 \times 10^{14},\; 1.5 \times 10^{13}\right)\ {\rm g}$ for \( \gamma = (0.3,\; 0.39) \), respectively. In summary, the inclusion of absorption induces an
\(\mathcal{O}(10)\) correction for \( \gamma = 0.3 \) and an \(\mathcal{O}(10^2)\) correction for
\( \gamma = 0.39 \).

\subsection{The low frequency regime}

In our previous analysis, we supposed that the high-frequency (optical limit) regime 
(\ref{Eq: high frequency}) is valid during all the absorption process. We can then calculate what is the condition for the low-frequency regime to be efficient. One should consider it if the wavelength of the radiation $\lambda_{lf} \sim \frac{2 \pi}{T}$ is of the order of the Schwarzschild radius $r_s$ given by Eq.~(\ref{Eq:rs}).
Applying the condition at formation time, $\frac{2 \pi}{\Tin}\geq r_s$, with $\Tin$
given by Eq.~(\ref{Eq:tin}), we obtain

\beq
\Min \lesssim \sqrt{\frac{\alpha}{3}}\frac{16 \pi^3 M_P}{\gamma}\simeq \frac{7.2\times 10^{-3}}{\gamma}~g\,,
\eeq

\noi
which never happens in the parameter space we are interested in. Only for very low collapse efficiency $\gamma\ll10^{-2}$, the mass of the PBH at formation is suficiently
small to possess a Schwarzschild radius of the order of the frequency of the plasma at formation time.

After formation time, the temperature of the plasma decreasing with time, the condition for a transition high--frequency $\rightarrow$ low--frequency becomes

\beq
\frac{T}{2 \pi}\lesssim \frac{4 \pi M_P^2}{R_\gamma \Min}~~\Rightarrow ~~
\frac{a}{\ain} \gtrsim \frac{R_\gamma}{8 \pi^2} \frac{\Min \Tin}{M_P^2} \,,
\eeq

\noi
or, using Eq.~(\ref{Eq:tin})

\beq
\frac{a}{\ain}\gtrsim\frac{R_\gamma}{8 \pi^2}\sqrt{\frac{\Min}{M_P}}
\left(\frac{4 \pi \gamma}{\sqrt{\alpha/3}}\right)^\frac12\simeq 
194\, R_\gamma\sqrt{\frac{\gamma}{0.3}}\sqrt{\frac{1~\rm g}{\Min}}\,.
\eeq

\noi
Therefore, there is always a region where low-frequency absorption dominates. However, at the initial stage after formation, high-frequency absorption is dominant. We have checked numerically that after the transition from the high- to low-frequency regime, there is no effect on the mass growth. For the analytical estimates, we can safely ignore the low-frequency absorption throughout the analysis.

As a conclusion, we have seen that the absorption of the plasma can have a strong impact for the evolution of PBH. Indeed, 
there exists a critical value for $\gamma$, given by Eq.~(\ref{Eq:deltac})
$\gamma_c\simeq0.395$, above which the absorption always dominates over the dilution. This also affects the evaporation process due to the mass gained by the absorption as one can see in Fig.~(\ref{Fig:m(a)}). It becomes then interesting to
study the indirect effect of the absorption to the dark matter production.

\section{Improved predictions for PBH reheating and the DM from PBHs}
\label{Sec:reheating}

\subsection{Consequences of the absorption on the reheating}

Until now, we have not considered the possibility that PBHs could also have an impact on primordial plasma. However, if they are present in sufficient quantities, they can affect both the reheating phase and the DM population through their evaporation. They may even be candidates for DM themselves if they are sufficiently stable.

Indeed, if the initial PBH abundance $\beta$ defined by

\beq
\beta=\frac{\rho_{BH}(\ain)}{\rho_{\rm tot}(\ain)}=\frac{n_{BH}(\ain)\Min}{\rho_{\rm tot}(\ain)}\,,
\label{Eq:beta}
\eeq

\noi
exceeds a certain critical threshold, $\beta_{\rm c}$, the Universe enters a PBH-dominated phase before their evaporation. In this scenario, known as PBH reheating
\cite{RiajulHaque:2023cqe}, the PBHs govern the cosmic energy budget until they evaporate, at which point their Hawking radiation repopulates the universe with Standard Model particles and possibly other species such as DM.
In the standard scenario, where thermal absorption is neglected, the critical $\beta$ value, $\beta_{\rm c}$ is given by~\cite{RiajulHaque:2023cqe, Haque:2023awl, Haque:2024eyh}
\begin{equation}
\beta_{\rm c} = \sqrt{\frac{3\, \epsilon}{8\pi\, \gamma}} \left( \frac{M_{\rm p}}{M_{\rm in}} \right)
\sim 7.3 \times 10^{-6} \left( \frac{1\, \mathrm{g}}{M_{\rm in}} \right) \left( \frac{0.2}{\gamma} \right)^{1/2},
\end{equation}
and the reheating temperature corresponds to the temperature at the end of PBH evaporation, can be written as

\beq
H(\arh)=\frac{2}{3\tev} =\sqrt{\frac{\alpha}{3}}\frac{\Trh^2}{M_P}\,,
\eeq

\noi
or, neglecting the thermal absorption  and using Eq.~(\ref{Eq:tev})

\begin{equation}
T_{\rm RH} = \left( \frac{360\, \epsilon^2}{\pi^2 g_\ast(T_{\rm RH})} \right)^{\frac{1}{4}}
\frac{M_{\rm p}^{\frac{5}{2}}}{M_{\rm in}^{\frac{3}{2}}}
\sim 3.6 \times 10^{10}\, \mathrm{GeV} \left( \frac{1\, \mathrm{g}}{M_{\rm in}} \right)^{\frac{3}{2}},
\label{Eq:trh}
\end{equation}

\noi
where we assume $g_\ast(T_{\rm RH}) = 106.75$.

However, the absorption affects the reheating.
Indeed, its major consequence is an increase of the mass of the PBH, therefore extending 
its lifetime. It becomes then clear that the reheating temperature which results from its 
evaporation is lower if we take absorption into account. Indeed, a PBH that evaporates later will release radiation into a larger Universe, thereby reducing $\rhorh$ and $\Trh$. More precisely, once thermal absorption is taken into account, both quantities
$\beta_{\rm c}$ and $\Trh$ receive corrections. 
From Eqs.~(\ref{Eq:beta}) and (\ref{Eq:trh}), they become:

\begin{equation}
\beta_{\rm c}^{\rm T} \sim R_\gamma^{-1} \beta_{\rm c}, \qquad
T_{\rm RH}^{\rm T} \sim R_\gamma^{-3/2}\, T_{\rm RH}\,.
\end{equation}

\noi
As an example, for a PBH of mass $1\, \mathrm{g}$ and $\gamma = 0.3$, the reheating temperature is reduced from $3.6 \times 10^{10}\, \mathrm{GeV}$ to $4.2 \times 10^{9}\, \mathrm{GeV}$ due to thermal absorption effects.\\

\subsection{Consequences of the absorption on dark matter production}

A second interesting consequence is the impact of the absorption phenomenon on DM production, through the evaporation of PBHs. We will focus on to the PBH reheating scenario with $\beta > \beta_{\rm c}$\footnote{We do not discuss the case $\beta < \beta_{\rm c}$, where PBHs evaporate 
before dominating the Universe. This choice is motivated by the fact that, in the PBH-
dominated scenario, the DM abundance becomes independent of $\beta$, reducing the 
parameter dependence and making the effect of thermal absorption more transparent.}. The 
key quantity in this context is the number of DM particles produced by a single PBH before it evaporates. Due to thermal absorption, the PBH mass increases, and this 
modifies the particle yield. As before, we assume that the mass growth occurs 
instantaneously just after formation, since the timescale associated with absorption is 
negligible compared to the total evaporation time. The number of particles emitted per PBH \cite{Haque:2023awl}

\beq
{\cal N}_i=\int_{\tin}^{\tev} \frac{dN_j}{dt} dt
\eeq

\noi
is modified to
\begin{equation}
    \mathcal{N}_i = \frac{15\,C\,\zeta(3)}{g_*(T_\text{BH})\,\pi^4}
    \begin{cases}
       \left(\frac{R_\gamma \,M_{\rm in}}{M_{\rm p}}\right)^2 \,,~m_j < T_\text{BH,\,T}^\text{in}\,,\\[8pt]
        \left(\frac{M_{P}}{m_j}\right)^2 \,,~m_j > T_\text{BH,\,T}^\text{in}\,,
    \end{cases}\label{eq:pbh-num}
\end{equation}

\noi
where $m_j$ is the mass of the dark matter species, and $C = (1,\, 3/4)$ for scalars and fermions, respectively, and where the PBH temperature after mass growth is

\begin{align}
& T_{\rm BH,\,T}^{\rm in}=R_\gamma^{-1} \frac{M_{P}^2}{\Min}
\simeq 10^{13}\left(\frac{1~\rm{g}}{R_\gamma\, M_{\rm in}}\right)~\rm{GeV}\,.  
\end{align}

The present-day DM relic abundance at temperature $T_0$ is given by~\cite{Mambrini:2021cwd}
\be
\Omega_jh^2= 1.6\times 10^8\,\frac{g_0}{g_{\rm RH}}\frac{{\cal N}_j\times n_{\rm BH}(t_{\rm ev})}{T_{\rm RH}^3}\,\frac{m_j}{\text{GeV}}\,,
\label{Eq:omegah2}  
\ee
where $n_{\rm BH}(\tev)$ is the PBH number density at evaporation,
and is assumed to follow instantaneous evaporation \cite{Haque:2023awl}

\beq
n_{\rm BH}(\tev)=\frac{\rho_{\rm BH}(\ain)}{\Min}\left(\frac{\ain}{\aev}\right)^3\,,
\eeq

\noi
or combining Eqs.~(\ref{Eq:min}), (\ref{Eq:beta}) and (\ref{Eq:xev}),

\begin{equation}
n_{\rm BH}(t_{\rm ev}) = 12\,\epsilon^2 \, \frac{M_{\rm p}^{10}}{\left(R_\gamma\,M_{\rm in}\right)^7}\,.
\end{equation}
The final DM relic density then takes the form~\cite{Haque:2024cdh}\footnote{For the case $m_j < T_{\rm BH,\,T}^{\rm in}$, constraints from warm dark matter (WDM) apply, as PBHs may emit highly energetic DM particles that remain relativistic until structure formation, conflicting with observations~\cite{Haque:2023awl, Barman:2024iht}.}:
\begin{equation}
   \frac{\Omega_jh^2}{0.12} =
    \begin{cases}
     \sqrt{\frac{1.14\times 10^8}{R_\gamma\,M_{\rm in}}}  \left(\frac{m_j}{\rm GeV}\right) \,,~m_j < T_\text{BH,\,T}^\text{in}\,,\\[8pt]
        \left(\frac{10^8\rm g}{R_\gamma\,M_{\rm in}}\right)^{\frac{5}{2}}\left(\frac{1.2\times 10^{10}\,\rm GeV}{m_j}\right) \,,~m_j > T_\text{BH,\,T}^\text{in}\,.
    \end{cases}\label{eq:pbh-relic}
\end{equation}

The main consequence of thermal absorption is an extension of the lifetime of PBH, which 
results in the injection of dark matter much later, and therefore greater dilution. This 
dilution, represented by $\xev$, is therefore greater for large values of $\gamma$. In 
other words, the parameter space allowed by the relic abundance constraints is more open, allowing for larger masses $m_j$, if absorption effects are taken into account.  

\begin{figure}[!ht]
\centering
\includegraphics[width = 0.45\textwidth]{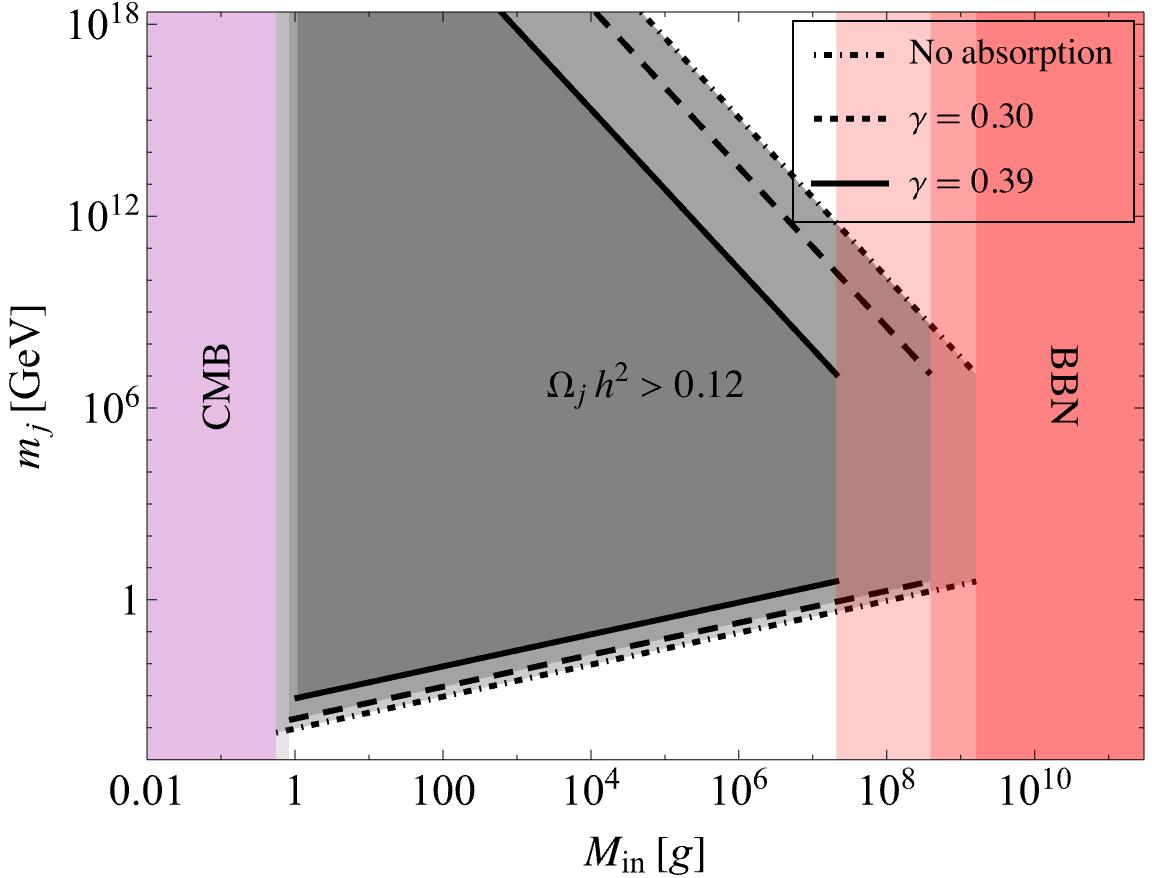}
\caption{Viable parameter space in the $(M_{\rm in}, m_j)$ plane for $\beta > \beta_c$. The black contours indicate regions consistent with the observed DM relic abundance. The dashed and solid lines correspond to $\gamma = 0.3$ and $\gamma = 0.39$, respectively in the presence of PBH mass
growth due to radiation absorption. For comparison, the standard scenario without absorption is also shown by the dot-dashed black line. Shaded regions are excluded due to DM overproduction, while the white regions are viable but require an additional production mechanism to match the observed relic abundance. BBN and CMB constraints exclude the red and magenta-shaded areas, respectively. The CMB bounds are derived from the horizon size at the end of inflation, assuming a de Sitter-like expansion, and incorporate the upper limit on the Hubble parameter set by the tensor-to-scalar ratio $r < 0.036$, as constrained by Planck 2018 and BICEP2/Keck Array observations~\cite{BICEP:2021xfz, BICEP2:2015nss}.}
\label{Fig: DM-evaporation}
\end{figure}

Fig.~\eqref{Fig: DM-evaporation} illustrates how the DM parameter space in the $(M_{\rm in},\, m_j)$ plane is modified when thermal absorption is taken into account. For $\gamma = 0.39$, we observe an $\mathcal{O}(10)$ correction in the region where $m_j < T_{\rm BH,\,T}^{\rm in}$, while for $m_j > T_{\rm BH,\,T}^{\rm in}$, the correction 
becomes more significant, reaching up to $\mathcal{O}(10^4)$. This shift in the high--mass region is indeed much more significant, since it concerns the region that is 
exponentially suppressed by the Boltzmann factor, which depends on the temperature of 
the PBH, itself greatly affected by thermal absorption. 
This shift is clearly visible in the figure and highlights the substantial impact of 
thermal mass growth on DM production from PBH evaporation.\\

\subsection{Consequences of the absorption on PBH relics}

Another interesting option is the case where the PBHs are sufficiently stable to 
constitute the DM content themselves. The effect of absorption also significantly changes the parameter space studied so far. 
Indeed, incorporating thermal absorption into PBH evolution significantly extends 
their lifetimes, thereby modifying the mass threshold for PBHs that can survive until 
today, as discussed in the previous section. This correction introduces a revised lower 
bound on the PBH mass that can survive today and naturally shifts the viable parameter 
space in which PBHs can constitute DM. The contribution of PBHs to the present dark matter density is quantified by the fraction
$f_{\rm PBH} = \frac{\Omega_{\rm PBH}}{\Omega_{\rm DM}}$,
where $\Omega_{\rm PBH}$ and $\Omega_{\rm DM}$ denote the present density parameters of 
PBHs and DM, respectively. Constraints on $f_{\rm PBH}(M_{\rm BH})$ arise from a variety 
of observational probes~\cite{Oncins:2022ydg, Carr:2016drx, Green:2020jor, Carr:2020xqk}, including gravitational microlensing~\cite{Niikura:2019kqi, Niikura:2017zjd}, extragalactic gamma-ray backgrounds~\cite{Carr:2016hva}, and 
gravitational wave observations~\cite{Nitz:2021vqh, Kavanagh:2018ggo}.

\begin{figure}[!ht]
\centering
\includegraphics[width = 0.45\textwidth]{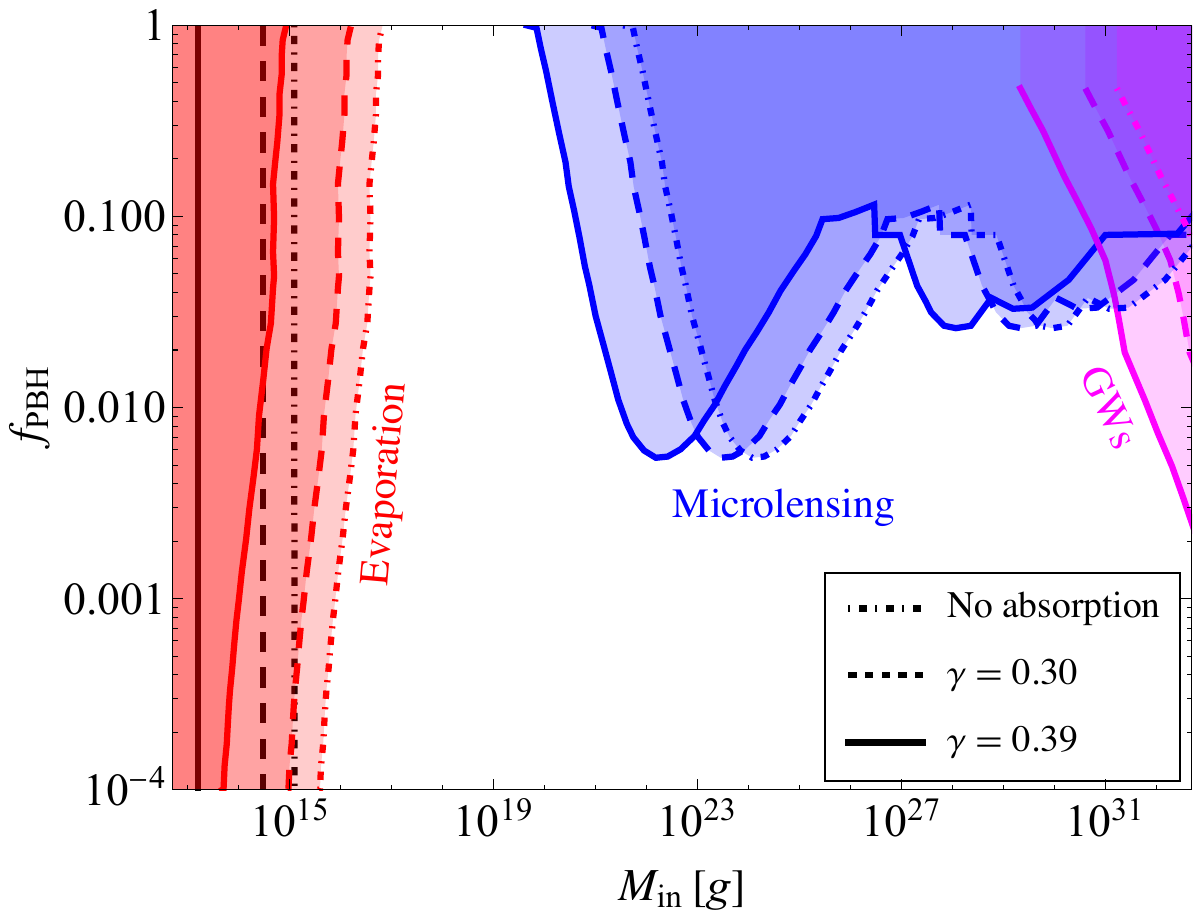}
\caption{Constraints on the PBH dark matter fraction $f_{\rm PBH}$ as a function of the initial mass $M_{\rm in}$. The dot-dashed curve represents the scenario with Hawking evaporation only, while the solid and dashed curves include the effect of mass growth due to thermal absorption for $\gamma = 0.30$ and $\gamma = 0.39$, respectively. The shaded regions indicate excluded parameter space from various observations: evaporation (red), microlensing (blue), and gravitational waves (magenta). The black lines denote the minimum PBH mass above which they remain stable today.}
\label{Fig: DM-stable}
\end{figure}

Since thermal 
absorption modifies the mass evolution of PBHs, it leads to a discernible shift in the 
allowed parameter space. As the effect of absorption is an extension of lifetime, viable 
regions open up for smaller values of $\Min$. As the effect of absorption is an 
extension of lifetime, viable regions open up for smaller values of $\Min$. And this 
effect is all the stronger as lifetime is extended, i.e. for larger values of $\gamma$.
This effect is illustrated in Fig.~\ref{Fig: DM-stable}, where the solid and dashed 
curves represent the modified bounds for $\gamma = 0.39$ and $\gamma = 0.30$, 
respectively, while the dot-dashed line corresponds to the standard scenario considering 
only Hawking evaporation. For $\gamma = 0.3$, the allowed PBH formation mass range 
expands to $10^{16}\,{\rm g} \lesssim M_{\rm in} \lesssim 10^{21}\,{\rm g}$, and for $\gamma = 0.39$, $5\times10^{14}\,{\rm g} \lesssim M_{\rm in} \lesssim 5\times10^{19}\rm g$, where PBHs 
can be treated entirely as DM. In contrast, the standard Hawking-only scenario permits a 
range of $5 \times 10^{16}\,{\rm g} \lesssim M_{\rm in} \lesssim 5 \times 10^{21}\,{\rm g}$. These results highlight how thermal absorption effects can substantially shift  
the viable mass window for PBHs as DM towards lower formation mass.

\section{Conclusion}
\label{Sec:conclusion}

In this work, we have investigated the evolution of PBHs in the early universe by 
accounting for the absorption of the surrounding thermal radiation bath. Incorporating 
radiation absorption into PBH dynamics leads to significant mass growth, which in turn extends their lifetimes and alters the standard predictions based solely on Hawking 
evaporation. For example, with $\gamma=0.30$, we find a 4 times increase in PBH mass and 
an associated $\mathcal{O}(10^2)$ enhancement in the lifetime compared to the standard Hawking only case. We also identified the critical conditions under which mass growth 
saturates, and found a critical value of the collapse efficiency parameter, $\gamma_{\rm c} \simeq 0.395$, above which PBHs undergo runaway mass growth.\\
These thermal corrections have notable cosmological consequences. They reduce the 
reheating temperature and shift the critical abundance $\beta_c$ required for PBH 
domination. In the context of DM, thermal absorption modifies the particle yield from 
PBH evaporation and lowers the minimum mass threshold for PBHs to survive until today, 
thereby shifting the viable parameter space in which PBHs can account for the entire DM 
abundance. For $\gamma = 0.39$, depending on the PBH formation mass, we find $\mathcal{O}(10)$ to $\mathcal{O}(10^4)$ corrections to the DM parameter space in scenarios involving PBH reheating, where we assume DM production solely from the evaporating PBHs.\\
These findings establish thermal absorption as a crucial ingredient in precision PBH 
cosmology. To our knowledge, this is the first comprehensive phenomenological study to 
systematically incorporate thermal absorption into PBH evolution and quantify its 
implications for reheating dynamics and DM production. A natural extension of this work 
would be to explore the role of thermal absorption in spinning (Kerr) PBHs and its impact on superradiance and gravitational wave signatures.

\section*{ACKNOWLEDGEMENTS}

The authors would like to thank Debaprasad Maity, Mathieu Gross and Lucien Heurtier for
helpful discussions. RK wants to thank the research group of Xian-Hui Ge at Shanghai University for various insightful discussions. 
This project has received funding from the European Union’s Horizon Europe research and innovation programme under the Marie Skłodowska-Curie Staff Exchange grant agreement No 101086085 – ASYMMETRY and the CNRS-IRP project UCMN.
We also acknowledge support by Institut Pascal at Université 
Paris-Saclay during the Paris-Saclay Astroparticle Symposium 2025.

\appendix
\section{High frequency absorption cross section and optical limit}
We begin by considering the Schwarzschild spacetime, described by the following line element in standard coordinates ($t,r,\theta,\phi$),
\be
\beal
ds^2=&\left(1-\frac{2G \Mbh}{r}\right)dt^2-\left(1-\frac{2G \Mbh}{r}\right)^{-1}dr^2\\
&~~~~~~~~~~-r^2d\theta^2-r^2\sin^2\theta d\phi^2.
\eeal
\ee

In the high-frequency limit, the absorption of massless particles can be approximated by the geodesic capture cross section, given by $\sigma_{hf}=\pi b^2_c$, where $b_c$ is the critical impact parameter \cite{Mambrini:2021cwd,Padmanabhan:2010zzb}. This parameter can be determined by analyzing the radial geodesic equation for Schwarzschild spacetime,
\be
\dot{r}^2+V_{eff}(r)=0
\ee
with 
\be
V_{eff}(r)=\left(1-\frac{2GM_{\rm BH}}{r}\right)\frac{\mathcal{L}^2}{r^2}-\mathcal{E}^2\,,
\label{Eq:veff}
\ee
where, $\mathcal{E}$ and $\mathcal{L}$ denote the conserved energy and angular momentum respectively. With this setup, the impact parameter is defined as, $b\equiv \mathcal{L}/\mathcal{E}$. 

The critical impact parameter corresponds to the trajectory of a massless particle that asymptotically approaches an unstable circular orbit, and is determined by the following conditions on the effective potential
\be
V_{eff}(r)|_{r=r_c}=0,~~\frac{dV_{eff}(r)}{dr}\Big|_{r=r_c}=0 \,,
\ee        
where, $r_c$ is the critical radius, corresponding to the radius of the unstable
circular orbit. Substituting the expression of the effective potential
(\ref{Eq:veff}), we derive the following equations \cite{Benone:2014qaa} 

\be
\frac{\mathcal{E}}{\mathcal{L}}=\pm \sqrt{\frac{1}{r^2_c}\left(1-\frac{2GM_{\rm BH}}{r_c}\right)},
\ee

\noi
and 
\be
r^4_c-3GM_{\rm BH}r^3_c=0\,.
\ee
The solution of the last equation determines the critical value of the radius for the unstable circular photon orbit as 
\beq
r_c=3GM_{\rm BH}=\frac32 r_s=\frac{3}{8 \pi}\frac{\Mbh}{M_P^2}\,.
\eeq

\noi
Therefore, the critical impact parameter can now be evaluated as 

\beq
\frac{1}{b_c^2}=\left(\frac{\mathcal{E}}{\mathcal{L}}\right)^2=\frac{1}{r_c^2}\left(1-\frac{r_s}{r_c}\right) \,,
\eeq

\noi 
or

\beq
b_c^2=\frac{27}{4}r_s^2\,.
\eeq

\noi
Accordingly the capture cross section then reads 

\beq
\sigma_{hf}=\pi b^2_c=\frac{27}{64 \pi} \frac{\Mbh^2}{M_P^4} \,.
\eeq

\section{Low frequency absorption cross section of Schwarzschild black hole}
In the following discussion, we will outline the steps to derive analytically the 
absorption cross section both for bosons and fermions in the low frequency limit. 
We will utilize the well known strategy for analytical derivation: matching the 
asymptotic solutions of different domains in the overlapping regions. The 
derivation will be very similar to what has been followed in \cite{Page:1976df} 
and \cite{Brito:2015oca}, however few changes have been made as the original analysis deals with Kerr BH. 

Notably, while the Teukolsky formalism is most commonly applied in Kerr 
spacetime, its Schwarzschild limit yields a simplified form suitable for studying spin-weighted fields in a spherically symmetric vacuum geometry. In the 
Schwarzschild background, the Teukolsky equation, governing the evolution of 
fields for generic spin $s$ (where $s$ represents the spin associated with 
scalar, photon, and spin-$1/2$ fermions for the consideration of the main text) can be expressed as

\be\label{master.eqn}
\beal
&\Delta^{-s+1}\frac{d}{dr}\left(\Delta^{s+1}\frac{dR}{dr}\right)\\
&~~~~~~+\left[\omega^2r^4-2iGM_{\rm BH}r^2\omega s+(2ir\omega s-\Lambda) \right]R=0\,.
\eeal
\ee

\noi
where $\Delta\equiv r(r-2GM_{\rm BH})$ and $\Lambda\equiv (l-s)(l+s+1)$. To analyse the asymptotic behavior of the radial solution it is convenient to rescale the independent variable as $x\equiv (r/2GM_{\rm BH})-1$ and rewrite the equation in the following manner:
\be\label{normalized.rad}
\beal
&x^2(x+1)^2\frac{d^2R}{dx^2}+(s+1)x(x+1)(2x+1)\frac{dR}{dx}\\
&+\left[k^2x^4+2iskx^3-\Lambda x(1+x)-isk(2x+1)+k^2\right]R=0\,,
\eeal
\ee
where $k\equiv 2GM_{\rm BH}\omega$. Now we study two different region in the domain of the radial solution. 

{\it Region-I}: This region is characterized by $kx\ll1$. Accordingly we approximate the radial equation as
\be
\beal
&x^2(x+1)^2\frac{d^2R}{dx^2}+(s+1)x(x+1)(2x+1)\frac{dR}{dx}\\
&+\left[k^2-\Lambda x(1+x)-isk(2x+1)\right]R=0\,.
\eeal
\ee

\noi
At this point, it is important to mention that the regularity of the solution in 
the near horizon requires that the solution should be ingoing near the horizon 
which falls within this region. A general solution satisfying this criteria, 
reads as

\be
R\sim A_1 x^{-s+ik}(x+1)^{-s-ik}F(\alpha,\beta, \gamma, -x)\,,
\ee

\noi
with $\alpha=-l-s, \beta=l-s+1$ and $\gamma=1-s-2ik$. Where $F$ stands for the Hypergeometric function \cite{abramowitz1964handbook}.
Expanding this solution for large $x$, we find
\be\label{soln.near}
R\sim A_1\left[x^{l-s}\frac{\Gamma(\gamma)\Gamma(\beta-\alpha)}{\Gamma(\beta)\Gamma(\gamma-\alpha)}+x^{-l-s-1}\frac{\Gamma(\gamma)\Gamma(\alpha-\beta)}{\Gamma(\alpha)\Gamma(\gamma-\beta)}\right]\,.
\ee
Again one can also directly obtain the radial solution at large $x$ from \eqref{normalized.rad}. This region is  denoted as far-region, consisting of the second region. 

{\it Region-II:} In the far region we approximate \eqref{normalized.rad} as 
\be
\frac{d^2R}{dx^2}+\frac{2(1+s)}{x}\frac{dR}{dx}+\left[k^2+\frac{2isk}{x}-\frac{\Lambda}{x^2}\right]R=0\,.
\ee

\noi
The general solution of this equation can be expressed as
\be
\beal
R=& C_1e^{-ikx}x^{l-s}U(l-s+1,2l+2,2ikx)\\
&~~~~C_2e^{-ikx}x^{-l-s-1}U(-l-s,-2l,2ikx)\,.
\eeal
\ee

\noi
To match with the solution of region-I, we approximate this solution 
for $kx\ll1$, which gives
\be
R\sim C_1 x^{l-s}+C_2 x^{-l-s-1}\,.
\ee

\noi
Comparing  with Eq.~\eqref{soln.near}, we obtain

\be
\beal
&C_1=A_1\frac{\Gamma(1-s+2ik)\Gamma(2l+1)}{\Gamma(l-s+1)\Gamma(l+1+2ik)},\\
&C_2=A_2\frac{\Gamma(1-s+2ik)\Gamma(-1-2l)}{\Gamma(-l+2ik)\Gamma(-l-s)}.
\eeal
\ee

\noi
Now, the general solution of Eq.~\eqref{master.eqn} near $r\to \infty$ can be expressed as a superposition of ingoing and outgoing wave, 
\be
R_{slm}\sim \mathcal{I}_s\frac{e^{-i\omega r}}{r}+\mathcal{R}_s\frac{e^{i\omega r}}{r^{2s+1}}\,,
\ee

\noi
where

\begin{widetext}
\be
\beal
\mathcal{I}_s &=\frac{1}{\omega}\Bigg[k^{l+1+s}\frac{C_2(-2i)^{l+s}\Gamma(-2l)}{\Gamma(-l+s)}+k^{s-l}\frac{C_1(-2i)^{s-l-1}\Gamma(2l+2)}{\Gamma(l+s+1)}\Bigg]\,,\\
\mathcal{R}_s &=\omega^{-2s-1}\Bigg[k^{l+1+s}\frac{C_2(2i)^{l-s}\Gamma(-2l)}{\Gamma(-l-s)}+k^{s-l}\frac{C_1(2i)^{-l-s-1}\Gamma(2l+2)}{\Gamma(l-s+1)}\Bigg]\,.\\
\eeal
\ee
\end{widetext}

In terms of the reflection amplitude, $\mathcal{R}_s$, 
and the incident amplitude, $\mathcal{I}_s$, we define the greybody factor as
\be
\Gamma_{slm}=1-\left|\frac{\mathcal{R}_s\mathcal{R}_{-s}}{\mathcal{I}_s\mathcal{I}_{-s}}\right|\,.
\ee
Substituting the expression of the amplitudes, and performing a bit of algebraic manipulation, the greybody factor for integer spin can be expressed as
\be
\Gamma_{slm}=16G^2M^2_{\rm BH}\omega^2(2GM_{\rm BH}\omega)^{2l}\left[\frac{(l-s)!(l+s)!}{(2l)!(2l+1)!!}\right]^2\,,
\ee
whereas for $1/2$-integer spin the greybody factor reads as
\be
\Gamma_{slm}=(2GM_{\rm BH}\omega)^{2l+1}\left[\frac{(l-s)!(l+s)!}{(2l)!(2l+1)!!}\right]^2\,.
\ee

Finally, the absorption cross section can simply be obtained using 
\be
\sigma\sim \frac{\pi(2l+1)}{\omega^2}\Gamma\,.
\ee

\noi
The factor involving the angular momentum mode $l$ arises from the spherical symmetry of the Schwarzschild spacetime, while the other factor results from the 
normalization of the incoming wave
to unit amplitude. Note that the maximum contribution in the total absorption cross section, in the low frequency limit, comes from $l=s$ mode. Such as for 
scalar, photon and fermion the absorption cross section determined by the 
dominant ($l=s$) mode are
\be
\sigma_{\rm lf}\sim
\left\{
\beal
&16\pi G^2M_{\rm BH}^2\,,~~~~~~~l=s=0,~{\rm scalar}\,,\\
&2\pi G^2M_{\rm BH}^2\,,~~~~~~~~~~l=s=1/2,~{\rm fermion}\,,\\
&\frac{64\pi}{3}G^4M_{\rm BH}^4\,\omega^2\,,~~~~~l=s=1,~ {\rm photon}\,.\\
\eeal\right.
\ee

\bibliography{Ref} 

 \end{document}